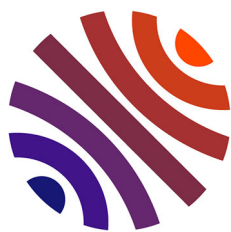

# HAL open science

# Long-term follow-up of DYT1 dystonia patients treated by deep brain stimulation: An open-label study

Laura Cif, Xavier Vasques, Victoria Gonzalez, Patrice Ravel, Brigitte Biolsi, Gwenaelle Collod-Beroud, Sylvie Tuffery-Giraud, Hassan Elfertit, Mireille Claustres, Philippe Coubes

## To cite this version:

## HAL Id: hal-01669958
### https://hal.science/hal-01669958v1

Submitted on 21 Dec 2017

# Long-Term Follow-Up of DYT1 Dystonia Patients Treated by Deep Brain Stimulation: An Open-Label Study

Laura Cif, MD,[1,2,3,4]* Xavier Vasques, PhD,[2,3,4] Victoria Gonzalez, MD,[1,2,3,4] Patrice Ravel, PhD,[3,5,6,7] Brigitte Biolsi, MD,[1] Gwenaelle Collod-Beroud, PhD,[3,8] Sylvie Tuffery-Giraud, PhD,[3,8] Hassan Elfertit, MD,[1] Mireille Claustres, MD, PhD,[3,8] and Philippe Coubes, MD, PhD,[1,2,3]

[1]*CHRU Montpellier, Hôpital Gui de Chauliac, Service de Neurochirurgie, Montpellier, France*
[2]*INSERM, U661, Montpellier, France*
[3]*Université de Montpellier 1, Montpellier, France*
[4]*CNRS UMR5203, Institut de Génomique Fonctionnelle, Montpellier, France*
[5]*CNRS UMR5048, Centre de Biochimie Structurale, Montpellier, France*
[6]*INSERM, U554, Montpellier, France*
[7]*Université de Montpellier 2, Montpellier, France*
[8]*INSERM, U827, Montpellier, France*

**Abstract:** Long-term efficacy of internal globus pallidus (GPi) deep-brain stimulation (DBS) in DYT1 dystonia and disease progression under DBS was studied. Twenty-six patients of this open-label study were divided into two groups: (A) with single bilateral GPi lead, (B) with a second bilateral GPi lead implanted owing to subsequent worsening of symptomatology. Dystonia was assessed with the Burke Scale. Appearance of new symptoms and distribution according to body region were recorded. In the whole cohort, significant decreases in motor and disability subscores ($P < 0.0001$) were observed at 1 year and maintained up to 10 years. Group B showed worsening of the symptoms. At 1 year, there were no significant differences between Groups A (without subsequent worsening) and B; at 5 years, a significant difference was found for motor and disability scores. Within Group B, four patients exhibited additional improvement after the second DBS surgery. In the 26 patients, significant difference ($P = 0.001$) was found between the number of body regions affected by dystonia preoperatively and over the whole follow-up. DBS efficacy in DYT1 dystonia can be maintained up to 10 years (two patients). New symptoms appear with long-term follow-up and may improve with additional leads in a subgroup of patients.

**Key words:** DYT1 dystonia; DBS; internal pallidum; disease progression

Primary DYT1 dystonia is an autosomal dominant movement disorder caused by a unique 3-base pair deletion (c.907delGAG) in the *TOR1A* gene on chromosome 9q34, leading to loss of a glutamate residue in the C-terminal region of the ATP-binding protein torsinA.[1] This mutation is a major cause of childhood-onset dystonia. Response to drug therapy is usually transient and unsatisfactory; botulinum toxin injections are rarely indicated.[2]

Following ablative surgeries in the treatment of movement disorders[3–5] and the advent of deep-brain stimulation (DBS) for treating Parkinson's disease and essential tremor[6,7] it has been known since 1999 that DBS of the internal globus pallidus (GPi)[8,9] is an effective symptomatic treatment for primary dystonic and hyperkinetic syndromes. Several groups have reported on the efficacy of GPi DBS for the treatment of primary dystonia.[10–13] Little is known about the very long-term results of DBS, the progression of the signs and the efficacy over time in the DYT1 population.

The primary aims of this study were to observe the durability of the initial response to pallidal stimulation and the evolution of the DYT1 motor phenotype during DBS. The study also aimed to establish whether the eventual worsening of symptoms despite DBS was due to underlying disease progression, loss of efficacy of DBS with time, or to a combination of both factors.

Long-term follow-up of DBS effects in dystonia patients is even more important than in other groups treated by DBS since it concerns patients who are much younger and therefore need to be followed over decades.

## PATIENTS AND METHODS

### Study Design

This is an open-label study conducted in the Academic Centre of Montpellier, France. Patients tested positive for the *DYT1* gene mutation[14] received clinical assessment and were videotaped according to a standardized protocol preoperatively and then at 6 and 12 months postoperatively and every year thereafter up to 10 years. Clinical assessments were carried out by at least two physicians trained in movement disorders (L.C., B.B., and P.C.) by using the two sections of the Burke-Fahn-Marsden Dystonia Rating Scale (BFMDRS).[15]

Twenty-six consecutive patients (16 female) were recruited from November 1997 to April 2004. The inclusion criteria were as follows: genetically confirmed DYT1 mutation; segmental or generalized dystonia; follow-up post-DBS of at least 3 years; normal neurological examination except for dystonia; absence of severe psychiatric disorders or other comorbidity increasing the surgical risks or compromising the clinical follow-up; lack of response to pharmacological treatments.

The whole population was first studied with bilateral single GPi electrode DBS. The mean preoperative scores (motor and disability) were compared with the mean worst scores with DBS to check for the response to stimulation. The mean follow-up at which the patients reached the best as well the worst scores were recorded. Scores at 3, 4, and 6 years were compared with scores at 1 year to establish whether a loss of efficacy occurred over time. From 7 to 10 years the population was too small to allow relevant statistical analysis. Nevertheless, individual evolution for each patient followed longer than 6 years is shown in Figure 2.

Due either to a subsequent worsening of dystonia or because of incomplete initial therapeutic effect (defined as less than 50% improvement in the motor scores), subsequently, patients were divided into Groups A and B according to whether they had received a second pair of bilateral GPi electrodes or not. Group A (18 patients) had a single pair of electrodes in the GPi during the whole follow-up, meanwhile Group B (8 patients) received a second pair of GPi electrodes. The second pair of electrodes occurred after at least 1 year of follow-up with the first pallidal electrodes, to reach steady state on these electrodes. According to the response to the second implantation within Group B, patients were divided in Groups B1 (significant additional improvement) and B2 (without significant additional improvement).

According to the motor subscale of the BFMDRS, nine body regions were identified and named by letters from A to I as follows: A, eyes; B, mouth; C, speech and swallowing; D, neck; E, right upper limb; F, left upper limb; G, trunk; H, right lower limb; and I, left lower limb. The total number of body regions affected preoperatively was calculated. Postoperatively, each time a new body part was affected by dystonia with the stimulation "ON," it was added to the number of regions affected preoperatively. The number of body regions involved by the disease at the time of the first surgery and at last follow-up have been compared (Table 1).

The frequency and severity of adverse events were also monitored.

Medication was modified during the second year of follow-up, in order not to interfere with the initial response to DBS. All the patients provided written informed consent for surgery.

### Surgical Procedure

Quadripolar electrodes (model 3389, Medtronic) were implanted bilaterally in the posteroventral GPi[16] in all 26 patients in a single procedure under general anesthesia. Additional electrodes were implanted in the GPi of eight patients (Group B). The targets were chosen by direct visualization on MRI, and confirmed immediately afterward by postoperative stereotactic MRI while still under general anesthesia. The electrodes were connected to the neurostimulators (Itrel 2, Itrel 3, Soletra or Kinetra, Medtronic) within 5 days of electrode implantation.

### Electrical Settings

Patients received stimulation in monopolar or bipolar mode with the following parameters: frequency 130 Hz; pulse width 450 μsec; amplitude between 0.3 and 2.1 V, according to the clinical response and the mode of stimulation (Table 2).

TABLE 1. *The 26 patients characteristics and treatments*

| Case | Gender | Age at onset (yr) | Age at first DBS surgery (yr) | Length of follow-up with DBS (yr) | Region of onset | Electrode addition | Position of the additional electrodes relative to the first implanted electrodes within the GPi | Treatment before surgery | Treatment at last follow-up | Number of regions before surgery | Regions involved before surgery | New regions during DBS | Number of new regions |
|---|---|---|---|---|---|---|---|---|---|---|---|---|---|
| 1 | F | 7,5 | 10,5 | 10 | I | – | | b,c,h | – | 7 | C,D,E,F,G,H,I | – | 0 |
| 2 | F | 6 | 26 | 10 | I | – | | b,c,h | g | 8 | B,C,D,E,F,G,H,I | A | 1 |
| 3 | F | 7 | 14 | 9 | E | – | | b,c,d,e | – | 7 | C,D,E,F,G,H,I | B | 1 |
| 4 | M | 7 | 8,5 | 8 | E | + | R posterior L posterior | c,d,f | – | 6 | D,E,F,G,H,I | B,C | 2 |
| 5 | M | 8 | 17 | 8 | E | – | | a,b,d,e | b,c,g | 6 | B,C,D,E,F,G | A,H,I | 3 |
| 6 | M | 9 | 13 | 7,5 | H | – | | c,h | – | 6 | D,E,F,G,H,I | – | 0 |
| 7 | F | 11,5 | 13,5 | 7,5 | I | – | | a,d,h | c,h | 2 | D,G | E,H | 2 |
| 8 | F | 9 | 13,5 | 7,5 | E | + | R posterior L posterior | c | – | 7 | A,B,D,E,F,G,H | – | 0 |
| 9 | M | 8 | 13 | 7 | I | + | R anterior L anterior | a,c,d | – | 4 | E,G,H,I | D | 1 |
| 10 | M | 7 | 13 | 7 | H | + | R posterior L anterior | a,d | b,c | 6 | C,D,E,G,H,I | F | 1 |
| 11 | M | 6,5 | 41,5 | 7 | H | – | | – | – | 7 | B,D,E,F,G,H,I | C | 1 |
| 12 | M | 6 | 13 | 7 | F | – | | d,h | – | 8 | A,B,C,E,F,G,H,I | – | 0 |
| 13 | F | 7 | 14 | 6,5 | I | + | R posterior L posterior | a,b,d,g | – | 6 | B,C,D,G,H,I | E | 1 |
| 14 | F | 7 | 8,5 | 6 | I | + | R posterior L posterior | a | – | 5 | E,F,G,H,I | – | 0 |
| 15 | F | 7 | 50 | 5,5 | I | – | | c,h | c,h | 7 | A,D,E,F,G,H,I | – | 0 |
| 16 | F | 20 | 25 | 5,5 | E | – | | c | c | 6 | D,E,F,G,H,I | – | 0 |
| 17 | F | 8 | 36,5 | 5,25 | E | – | | c | c | 7 | C,D,E,F,G,H,I | – | 0 |
| 18 | F | 6,5 | 19 | 5 | E | – | | d,e | – | 6 | D,E,F,G,H,I | – | 0 |
| 19 | F | 6,5 | 9,25 | 4,75 | H | + | R posterior L posterior | c,f,h | b | 8 | B,C,D,E,F,G,H,I | – | 0 |
| 20 | F | 9 | 65,5 | 4,25 | E | – | | c | c | 5 | D,E,G,H,I | F | 1 |
| 21 | M | 8 | 48 | 4 | I | – | | c | c | 7 | C,D,E,F,G,H,I | – | 0 |
| 22 | M | 8 | 35,25 | 4 | I | – | | a,c,f | c | 7 | C,D,E,F,G,H,I | – | 0 |
| 23 | F | 9 | 12 | 3,75 | I | – | | c | – | 4 | E,G,H,I | – | 0 |
| 24 | F | 6 | 9 | 3,5 | H | – | | d | – | 4 | E,G,H,I | – | 0 |
| 25 | F | 9 | 10,5 | 3 | D | – | | a,b,d | – | 7 | C,D,E,F,G,H,I | – | 0 |
| 26 | M | 6 | 23,5 | 6,5 | H | + | R anterior L anterior | c,h | c,h | 9 | A,B,C,D,E,F,G,H,I | – | 0 |

Dystonia distribution according to body parts is presented.
[a]L-dopa.
[b]Baclofen.
[c]Benzodiazepine.
[d]Anticholinergics.
[e]Terabenazine.
[f]Carbamazepine.
[g]Botulinum toxin.
[h]Others.
M = male; F= female; + = present; − = absent.

**TABLE 2.** *Motor scores of the 26 patients before the first DBS surgery, at best with a single pair of leads, before implantation of an additional pair of leads, and at last follow-up, as well as the corresponding electrical settings*

| Case | Motor scores (/120) before first DBS | Best motor (/120) scores | Motor scores (/120) before leads addition | Motor score at last follow-up | Settings at best before addition | Settings at last FU |
|---|---|---|---|---|---|---|
| 1 | 81 | 0 | No | 0 | R 1.7V E1-<br>L 1.7V E1- | R 1.7V E1-<br>L 1.8V E1- |
| 2 | 97,5 | 15 | No | 27 | R 1.8V E1-<br>L 1.5V E0-E1- | R 1.7V E1- E2-<br>L 1.5V E0-E1- |
| 3 | 63 | 1 | No | 5 | R1.8V E1-<br>L 1.5V E1- | R 1.7V E0-E1-E2-<br>L 1.4V E0-E1- |
| 4 | 76 | 0 | 30 | 30 | R 1.5V E1-<br><br>L 1.5V E1- | R (post)1.3VE1-;<br>R (ant) 1.0V E0-E1-<br>L (post) 1.4V E1-E2-;<br>L (ant) 2.0V E1-E2- |
| 5 | 35,5 | 25 | No | 29.5 | R 1.3V E1-<br>L 1.3V E1- | R 1.3V E1-E2-<br>L 1.7V E1-E2- |
| 6 | 51 | 0 | No | 15 | R 1.3V E1-E2-<br>L1.3V E1-E2- | R 1.5V E1-E2-<br>L 1.6V E1-E2- |
| 7 | 24 | 0 | No | 15.5 | R 1.8V E1-<br>L 1.7V E1- | R 1.5V E1-<br>L 1.1V E0-E1-E2- |
| 8 | 37,5 | 4 | 21 | 6 | R 2.0V E1-E2-<br><br>L 2.0V E1-E2- | R CH1 (post) 1.0VE1- E2-;<br>R CH2 (ant) 1.1V E1-<br>L CH1 (post) 1.0V E1-E2-;<br>L CH2 (ant) 1.0 V E1- |
| 9 | 37 | 0 | 16.5 | 4 | R 1.7V E0-E1-<br><br>L 1.7V E0-E1- | R CH1(ant) 1.9VE0-E1-;<br>CH2 (post) 1V E1-<br>L CH1 (ant) 1.5V E1-E2-;<br>CH2 (post) 1.4V E1-E2- |
| 10 | 65,5 | 9.5 | 30 | 8,5 | R 2.1V E1-E2-<br><br>L 1.4V E1-E2- | R (ant) CH1 1.2V E1-E2-;<br>R (post) CH2 0.9 V E1-E2-<br>L (ant) CH2 1.1V E1-E2-;<br>L (post) CH1 1.4V E1-E2- |
| 11 | 84,5 | 28 | No | 28 | R 1.5V E0-E1-E2-<br>L 1.5VE0-E1-E2- | R 1.6V E0-E1-E2-<br>L 1.4V E0-E2-E2- |
| 12 | 69 | 0 | No | 2 | R 1.6V E1-<br>L 1.5 V E2- | R 1.3V E2-<br>L 1.3V E1- |
| 13 | 46 | 28 | 29 | 35 | R 1.7V E0-E1-E2-<br><br>L 1.7V E0-E1-E2- | R CH1 (post) 1.3V E3-;<br>CH2 (ant) 1.4V E1-E2-<br>L CH1 (post) 1.5VE1-;<br>CH2 (ant) 1.0V E0-E1- |
| 14 | 35,5 | 0 | 31 | 0 | R 1.4V E1-<br><br>L 1.6V E1- | R CH1 (ant) 0.9V E0-E1-;<br>R CH2 (post) 0.8V E1-<br>L CH1 (ant) 1.0V E1-E2-;<br>L CH2 (post) 1.4V E1-E2- |
| 15 | 28 | 1 | No | 2 | R 1.8V E1-<br>L 1.8V E1- | R 1.3V E1-E2-<br>L 1.6V E1-E2- |
| 16 | 38 | 0 | No | 2 | R 1.1V E1-<br>L 1.1V E1- | R 1.4V E1-<br>L 1.1V E1-E2- |
| 17 | 67 | 9.5 | No | 16 | R 1.5 E1-E2-<br>L1.8V E0-E1- | R 1.5V E1-E2-<br>L 1.5V E0-E1-E2- |
| 18 | 64 | 0 | No | 0 | R 1.6 V E1-<br>G 1.5 V E1- | R 1.3V E1-<br>G 1.3V E1- |
| 19 | 101,5 | 48 | 57 | 54 | R 1.8V E1-E2-<br><br>L 1.8V E1-E2- | R CH1 (post) 1.0V E1-E2+E3-;<br>CH2 (ant) 1.3V E1-E2+<br>L CH1 (post) 1.3V E1-E2+;<br>CH2 (ant) 1.3V E1-E2+ |
| 20 | 27,5 | 9 | No | 10.5 | R 0.3V E1-<br>L 1.1V E1- | R 0.3 V E1-<br>L 1.1V E1- |
| 21 | 61,5 | 19 | No | 25 | R1.3V E2-E3-<br>L1.3V E2-E3- | R1.3V E2-E3-<br>L1.3V E2-E3- |
| 22 | 51 | 23 | No | 41 | R 0.8V E1-E2-<br>L 0.9V E1-E2- | R 1.3V E0+E1-E2-E3+<br>L 1.0V E0+E1+E2- |
| 23 | 40 | 0 | No | 5 | R 1.4V E1-<br>L 1.4V E1- | R 2V E1-<br>L 1.1V E1- |
| 24 | 42 | 3 | No | 3 | R 1.6V E1-E2-<br>L 1.5V E1- | R 1.6V E1-E2-<br>L 1.5V E1- |
| 25 | 60 | 0 | No | 0 | R 1.0V E1-<br>L 1.0V E1- | R 1.0V E1-<br>L 1.0V E1- |
| 26 | 112 | 46.5 | 53,5 | 69 | R 1.6V E1-E2-<br><br>L 1.6V E1-E2- | R CH1(ant) 0.9V E0-E1-;<br>CH2 (post) 1.2V E1-E2+<br>G CH1(ant) 1.4V E0-E1-;<br>CH2 (post) 1.2V E1-E2+ |

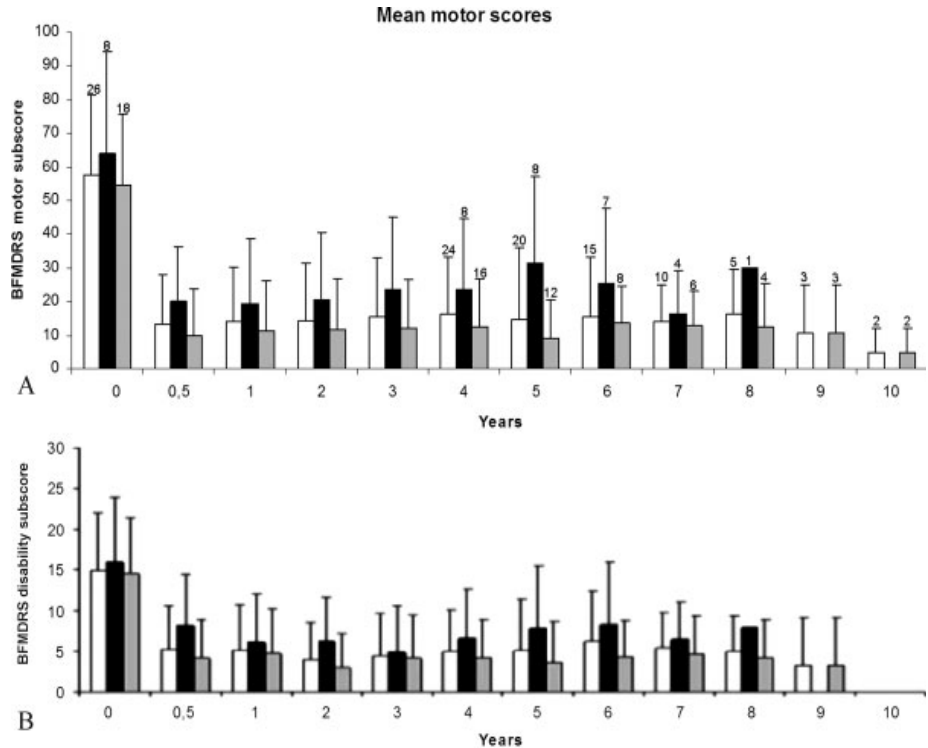

**FIG. 1.** BFMDRS motor and disability subscores evolution for the whole population (white plots), as well as for Group A (gray plots) and Group B (black plots). At 10 years, plots are missing because disability score is zero for the two patients.

### Statistical Analysis

Statistical analysis was performed with the R free software environment for statistical computing (Free software Foundation's GNU). The mean ± SD was used for quantitative variables. Discrete variables are presented as absolute numbers and percentages.

If the sample variance did not differ with an alpha of 0.05 ($F$ test for equality of variance) with a normal distribution (Shapiro-Wilk test), statistical comparisons of observed values were performed using a two-tailed $t$ test. Owing to the small sample size and the fact that some of the data were not normally distributed, comparisons were performed using a Mann-Whitney $U$ test for independent and continuous variables. Wilcoxon signed-rank tests, with corrections if necessary, were used for the matched and continuous variables and for the eight patients who received additional electrodes to illustrate the difference between the two subgroups, even if the population (subgroup B1: 4 patients, subgroup B2: 4 patients) was too small to allow relevant statistical analysis.

### RESULTS

The clinical characteristics of the patients and their treatments are shown in Table 1.

### Results of the Whole Population

For the 26 patients (16 children), the mean age at disease onset was 8.1 ± 2.8 years (6–20 years) and 21.6 ± 15.6 years (8.5–65.5 years) at surgery.

The mean follow-up period for the 26 patients was 6.2 ± 1.9 years (3–10 years).

### Comparison of Pre- and Postoperative Scores

The mean worst postoperative scores (motor: 23.3 ± 19.2, disability: 7.4 ± 6.7) at long-term follow-up were significantly lower ($P < 0.0001$ for both parts) than the mean preoperative scores (motor: 57.5 ± 23.9, disability: 14.9 ± 7.1).

### Motor and Disability Scores Evolution with DBS

Scores evolution with DBS for the 26 patients is presented in Figures 1A, B and 2A, B and in Table 3, which compares the mean scores at 3, 5, and 6 years with the mean preoperative scores.

The mean follow-up period when the best (motor and disability) scores were obtained was 1.7 ± 1.2 years and 1.9 ± 1.5 years, respectively and for the worst (motor and disability) scores 4.35 ± 2.1 years and 3.03 ± 2.3 years, respectively.

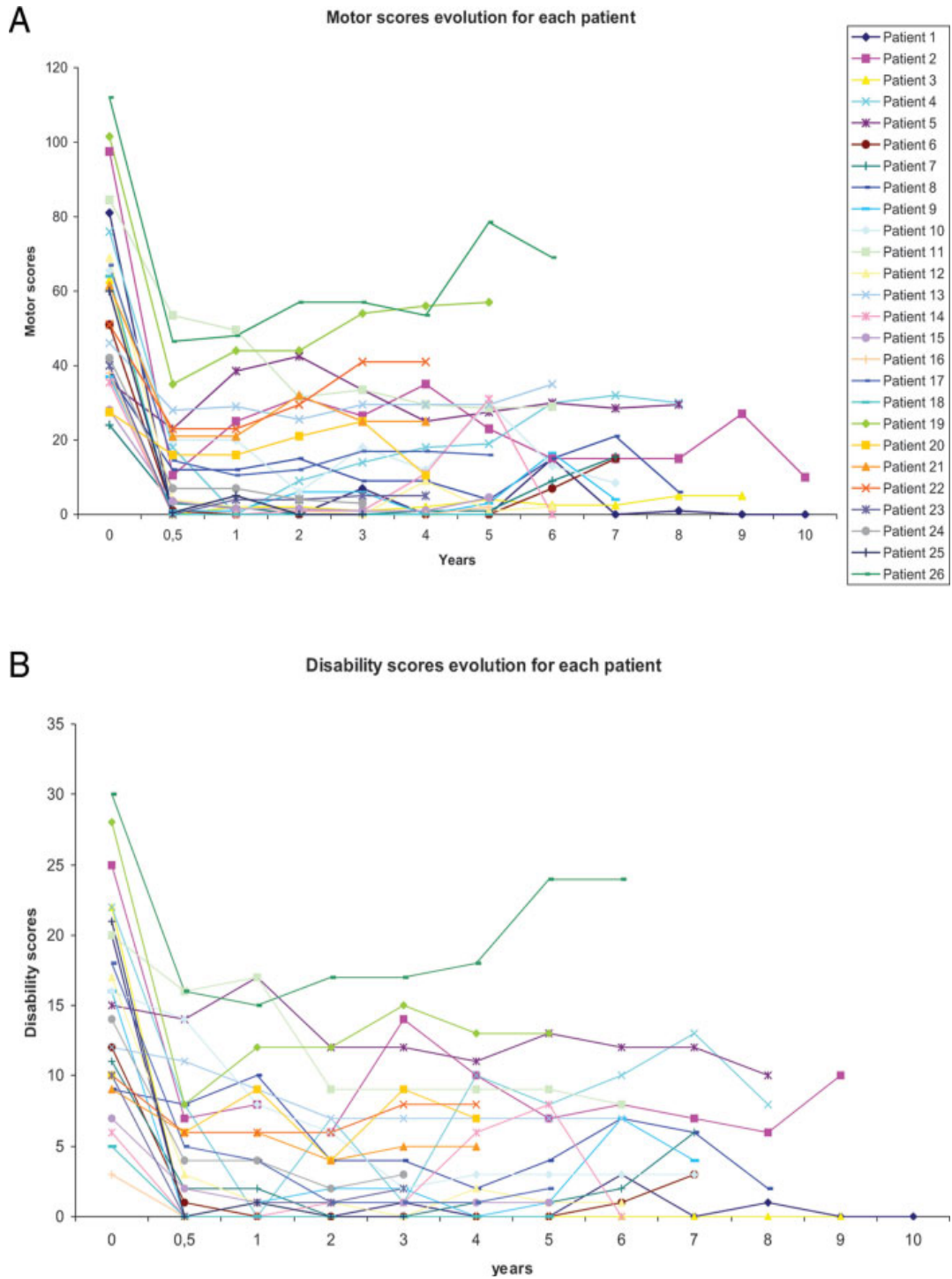

**FIG. 2.** (**A**) Motor scores evolution for each patient; (**B**) disability scores evolution for each patient; (**C**) evolution of motor scores in subgroup B1; (**D**) evolution of motor scores in subgroup B2.

Worsening of the scores (motor and disability) occurred in some individuals at more than 3 years follow-up, but comparison of the scores at different time points for the whole population showed that the efficacy of DBS did not decrease significantly over a period of up to 10 years after surgery.

Complete initial therapeutic response with persistence of the improvement over time has been observed

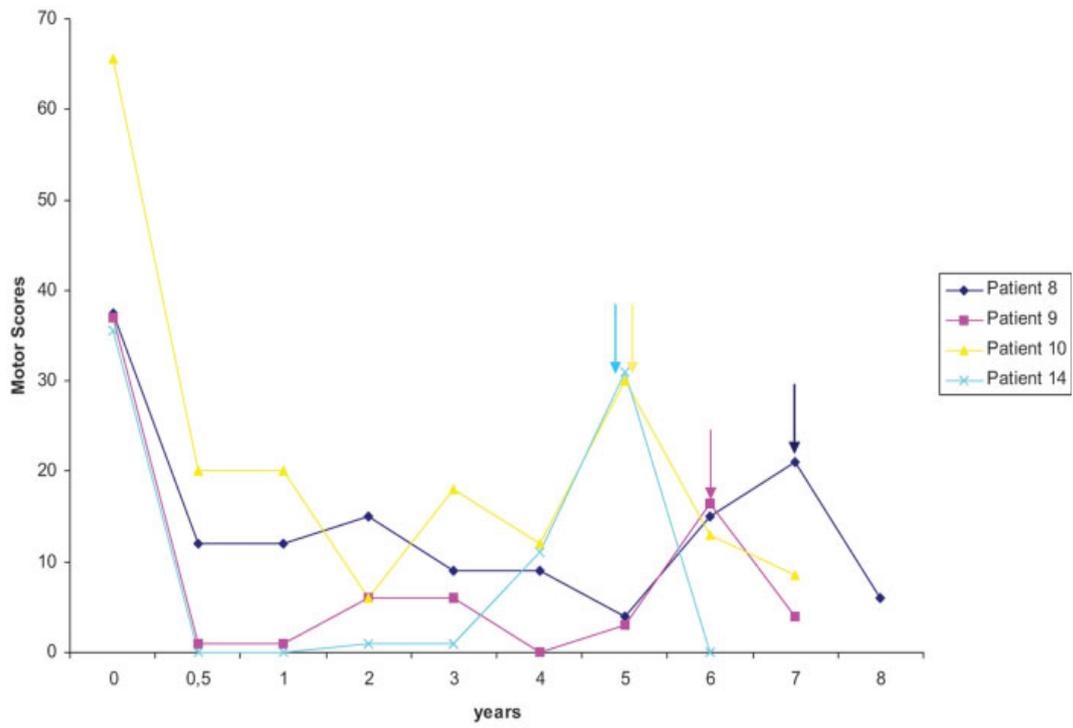
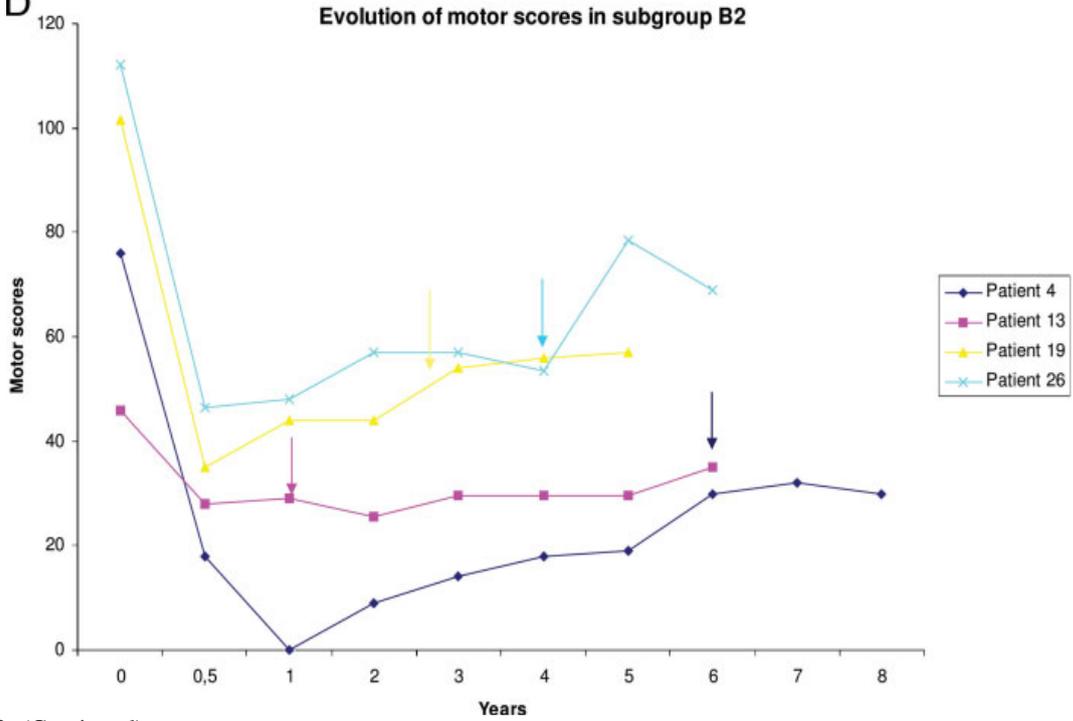

**FIG. 2.** (Continued)

**TABLE 3.** *Comparison (first two lines) between preoperative scores and 1, 3, 4, 5 and 6 yrs postoperative scores for the whole population. Lines 3 to 5 and 6 to 8 compare the motor and disability scores between Groups A and B*

|  | Before surgery | 1 yr after surgery | 3 yr after surgery | 4 yr after surgery | 5 yr after surgery | 6 yr after surgery |
|---|---|---|---|---|---|---|
| Motor scores of the 26 patients | 57.51 ± 23.94 | 13.88 ± 16.26 ($P < 0.0001$) | 15.65 ± 17.38 ($P < 0.0001$) | 16.25 ± 17.11 ($P < 0.0001$) | 14.78 ± 21.10 ($P = 0.001$) | 15.64 ± 17.63 ($P = 0.003$) |
| Disability scores of the 26 patients | 14.92 ± 7.11 | 5.11 ± 5.55 ($P < 0.0001$) | 4.38 ± 5.27 ($P < 0.0001$) | 4.95 ± 5.10 ($P < 0.0001$) | 5.1 ± 6.21 ($P = 0.001$) | 6.2 ± 6.22 ($P = 0.008$) |
| Group A motor scores | 54.69 ± 20.92 | 11.5 ± 14.58 | 12.13 ± 14.55 | 12.56 ± 14.17 | 8.95 ± 11.41 | 13.68 ± 10.90 |
| Comparison of the motor scores between Groups A and B | $P = 0.560$ | $P = 0.538$ | $P = 0.093$ | $P = 0.157$ | $P = 0.010$ | $P = 0.223$ |
| Group B motor score | 63.87 ± 30.29 | 19.25 ± 19.49 | 23.56 ± 21.48 | 23.62 ± 20.95 | 31.5 ± 25.60 | 25.5 ± 22.36 |
| Group A disability score | 14.57 ± 6.90 | 4.73 ± 5.54 | 4.21 ± 5.24 | 4.25 ± 4.66 | 3.61 ± 5.04 | 4.37 ± 4.37 |
| Comparison of the disability scores evolution between Groups A and B | $P = 0.367$ | $P = 0.338$ | $P = 0.128$ | $P = 0.117$ | $P = 0.021$ | $P = 0.415$ |
| Group B disability score | 15.85 ± 8.13 | 6.14 ± 5.87 | 4.85 ± 5.75 | 6.57 ± 6.05 | 7.85 ± 7.60 | 8.28 ± 7.65 |

(Patients 1, 12, 18, 24, and 25) as well as more limited initial responses (5, 13, 19, 21, 22, and 26). In other patients (3, 4, 6, 7, 9, 14, 16), after an initial complete therapeutic response, recurrence of previously controlled signs or the appearance of new signs also occurred. In accordance with these findings, the addition of a second pair of leads in several patients (8, 9, 10, and 14) led to an improvement in the therapeutic effect obtained by single bilateral GPi leads probably due to a somatotopic organization within the GPi.[17–19] For example, in Patient 8, the initial lead implanted was anteriorly placed and permitted control of the axial and lower limb dystonia only. The additional leads were implanted posterior to the first leads, and completely suppressed the residual dystonia involving the upper limbs. On the contrary, in patient 9, after an initial complete response, a second lead was implanted more anteriorly within the GPi in order to control the recurrence of the lower limb dystonia.

### Comparisons Between Group A and B

The mean of age at symptoms onset for group B was 7.2 ± 0.9 years and 8.4 ± 3 years for Group A ($P = 0.297$). The mean of age at first DBS surgery was 12.9 ± 4.9 years for Group B and of 25.5 ± 16.7 years for Group A ($P = 0.04$).

For group B, the mean follow-up with second DBS was of 2.16 ± 1.6 years.

The mean preoperative motor score for Group B was of 63 ± 30.3 and of 54.7 ± 20.9 for group A ($P = 0.567$).

The mean preoperative disability score for Group B was 17.4 ± 8.7 (6–30) and 13.8 ± 6.3 (3–25) for Group A ($P = 0.367$).

No significant differences were found between the two groups at 1, 3, and 4 years follow-up for both scores. At 5 years follow-up, significant differences were found between the two groups in motor ($P = 0.01$) and disability ($P = 0.025$) scores.

In Group B, the difference between the preoperative scores before first surgery and the scores before the second surgery remained significant for both motor ($P = 0.008$) and disability ($P = 0.016$) scores.

### Evolution with Additional Lead Implantation: Group B

At the time of the second pair of electrodes implantation the mean follow-up with DBS was 4.8 ± 1.9 years (1.5–7 years) and the mean of age was 17.7 ± 4.8 years (12–27 years). No significant differences were found for Group B between the scores before implantation of the additional electrode (motor: 33.5 ± 14.4; disability: 9.5 ± 4.9) and at last follow-up (motor: 25.8 ± 25.6; disability: 7.8 ± 7.7).

Patients could be classified into two subgroups according to the motor scores evolution after the second implantation: B1 (four patients) with significant improvement ($P = 0.017$) and B2 (four patients) without significant improvement ($P = 0.339$). Comparing the baseline scores (before the first operation) of the two subgroups, there was a trend for a significant difference ($P = 0.05$), showing that the group which improved after the implantation of the second electrode had less severe baseline scores than the group that did not. Nevertheless, no significant difference was recorded when comparing the preoperative scores of Groups A and B. This renders difficult to interpret the influence of the initial disease severity on the response to DBS. Comparing the response at 1 year after the first DBS surgery, no significant difference was recorded between the two subgroups ($P = 0.113$).

### Number of Regions Involved by the Disease; Does Disease Progression Occur with DBS?

Before surgery, 162 (6.2 ± 1.5) body parts were affected by dystonia for the whole cohort (Table 1). During the total follow-up period (pre- and postoperative), 176 (6.8 ± 1.5) body parts were affected by dystonia. The number of regions involved during the whole follow-up and independently on DBS therapy, was significantly higher than before DBS ($P = 0.001$) suggesting disease evolution.

### Adverse Events

There were no hemorrhagic complications. Hardware-related complications included two extension cable breakages (Patients 5 and 9) and one spontaneous early IPG dysfunction after 3 months (Patient 18). Hardware infections occurred in Patients 4 and 10.

The mean life span for the IPGs was 3.1 years.

## DISCUSSION

We studied the durability of the GPi DBS efficacy in 26 patients with primary DYT1 dystonia with a follow-up period of at least 3 years and up to 10 years. All the patients underwent the same surgical procedure, clinical assessment, and protocol for electrical settings, which allowed comparisons between patients.

The initial efficacy of GPi DBS in DYT1 and other primary dystonias has been evaluated previously but the reported follow-up did not exceed 3 years.[13,20] Loher et al.[21] reported follow-up longer than three years in a heterogeneous dystonia population of nine patients concluding that DBS maintains symptomatic and functional improvement with long-term follow-up. The younger age of our patients may partially explain the evolution of the disease with DBS. However, this study included only one late onset DYT1 dystonia patient with a follow-up of 5 years. Cersosimo et al.[22] reported on one DYT1 patient treated by GPi DBS and followed for more than 3 years with a limited overall improvement.

The heterogeneity of the effect of DBS in primary dystonia has been reported previously.[23] Several factors, including surgical protocols, contact position within the target,[24,25] DYT1 status,[9–11,24] age at surgery[11] but especially disease duration[26] could all be part of the explanation for the observed differences in results.

Notwithstanding the fact that the overall response pattern was very satisfactory, the evolution of dystonia with DBS was different from one patient to another. Patients with dystonic storm responded as well as patients in a stable clinical condition. The location of the electrodes was verified by the postoperative MRI confirming optimal lead placement. The severity of dystonia preoperatively did not influence the outcome of DBS. Nevertheless, patients without significant improvement with additional lead implantation (B2) exhibited more severe preoperative motor scores than patients with improvement following additional lead implantation (B1).

With follow-up extending up to 10 years, worsening of the motor and disability scores of the patients with DBS has been observed in the population of DYT1 dystonia, sometimes after even 5 years during which patients were free of symptoms. Symptoms occurred in the body parts previously involved, in new body parts, or in both.

Group A (higher age at surgery) did not require the additional leads meaning that worsening of symptoms occurs more rarely in older patients receiving DBS surgery. DBS early administered allows a better response but will not prevent from disease progression.

Numerous studies have focused on the pathophysiological mechanisms involving the mutated torsinA protein.[27] Nevertheless, little is known about the relationships between the genotype, pathophysiology and the broad phenotype of the disease. Identification of new interactors of torsinA as Printor and coworkers[28] opens new fields of investigation. Variations in the initial therapeutic response to DBS and sustained efficacy could be linked to the complex pathophysiological mechanisms of the disease with the implication of population specific genetic modifiers in incomplete penetrance or clinical variability.[29,30]

DBS has been described as being an adaptable therapy but in our experience long-term adaptability was limited. Increasing the number of activated contacts and/or the voltage does not always provide additional improvement and control of all the signs in patients who respond to the therapy.

An off-stimulation test would be of great interest to measure the actual extent of the disease several years after DBS surgery. In the present work, we certainly under-estimated the new signs of the disease that would occur, since patients were indeed benefiting from the DBS therapy. The delay of recurrence of symptoms in dystonia is variable from one patient to another, taking anything from a few minutes up to several months. This makes difficult the design of standardized off-stimulation protocols for the whole population.

## CONCLUSIONS

In primary DYT1 dystonia, GPi DBS efficacy is maintained up to ten years follow-up. The natural course of childhood onset DYT1 dystonia can be severe, with life-threatening complications. With DBS treatment, all the patients survived during the follow-up period, regardless of the severity of dystonia before surgery.

DYT1 dystonia seems to be a progressive disease even with DBS. With follow-up longer than 3 years, there is recurrence of signs or emergence of new signs, which cannot always be controlled by single lead implantation. In some patients, multiple electrodes within the GPi could be helpful in controlling the residual signs of the disease.

**Acknowledgments:** We are grateful to Professor Marwan Hariz, Queen's Square, The National Hospital for Neurology and Neurosurgery, London, for fruitful discussions.

Drs. Laura Cif, Xavier Vasques, Victoria Gonzalez, Patrice Ravel, Brigitte Biolsi and Hassan Elfertit have no financial or other disclosure to make.

Drs. Mireille Claustres, Gwenaelle Collod-Beroud, and Sylvie Tuffery-Giraud have no funding sources or potential conflicts of interest related to the study reported in this article. For the past year, Drs. Mireille Claustres, Gwenaelle Collod-Beroud and Doctor Sylvie Tuffery-Giraud had founding sources from Université Montpellier 1 and several patients associations (AFM, VLM, AMADYS-LFCD). For the past year, Doctor Philippe Coubes was consultant for Medtronic Company.

**Author Roles:** LC: Research Project (Conception, Organization, and Execution); Statistical analysis (Design and Execution); Manuscript (Writing of the first draft). XV: Research Project (Organization); Statistical analysis (Execution); Manuscript (Review and critique). VG: Research Project (Execution); Statistical analysis (Review and critique); Manuscript (Review and critique); PR: Research Project (Conception); Statistical analysis (Design); Manuscript (Review and critique). BB: Research Project (Execution); Statistical analysis (Review and critique); Manuscript (Review and critique). GC-B: Research Project (Organization); Statistical analysis (Review and critique); Manuscript (Review and critique). ST-G: Research project (Organization); Manuscript (Review and critique). HE: Manuscript (Review and critique). MC: Research Project (Organization). PC: Research Project (Conception); Statistical analysis (Review and critique); Manuscript (Review and critique).